# Turbo Interleaving inside the cdma2000 and W-CDMA Mobile Communication Systems: A Tutorial

Fabio G. Guerrero*

Electrical and Electronics Engineering School, Universidad del Valle, Cali, Colombia (South America).

Ciudad Universitaria Meléndez, Calle 13 100-00, Edificio 355, Oficina 205, Cali, Colombia

Phone: (572) 3392140 ext 109; Fax: (572) 3392140 ext 112

e-mail: fguerrer@univalle.edu.co

Maribell Sacanamboy

Department of Computer Science Engineering, Universidad Javeriana, Cali, Colombia (South America).

Calle 18 No. 118-250 Cali, Colombia

Phone: (572) 321 8200

e-mail: msacanamboy@puj.edu.co

(We would welcome considering this work for possible publication)

**Abstract**

In this paper a discussion of the detailed operation of the interleavers used by the turbo codes defined on the telecommunications standards cdma2000 (3GPP2 C.S0024-B V2.0) and W-CDMA (3GPP TS 25.212 V7.4.0) is presented. Differences in the approach used by each turbo interleaver as well as dispersion analysis and frequency analysis are also discussed. Two examples are presented to illustrate the complete interleaving process defined by each standard. These two interleaving approaches are also representative for other communications standards.

## 1. Introduction

Near one decade ago the IMT2000 initiative of the International Telecommunication Union identified five base station-mobile station air interfaces for the third-generation mobile communications systems (3G). It seems clear at the present however that the two technologies that will be dominating the global market of third generation mobile communications systems are cdma2000 and W-CDMA [1].

The salient feature of third generation mobile communications systems is its high capacity for transmitting information over the system data channels. By 1997 IMT2000 defined in Recommendation ITU-R M.1225 test data rates of 2048 kbit/s, 144 kbit/s and 64 kbit/s for indoors, pedestrian and vehicular traffic respectively [2] for purposes of evaluating the third-generation technologies. As expected, the continuing evolution of mobile technologies has left behind these values with much higher speeds. For instance, the Ultra Mobile Broadband™ (UMB™) air interface specification [3] is intended to deliver downlink and uplink data rates of 288 Mbit/s and 75 Mbit/s respectively using a bandwidth of 20 MHz.

To offer these high data rates with access terminals increasingly both small and functional it is imperative to work at the limit of efficiency in data transmission.

As it is well known in 1948 C. E. Shannon proved that the fundamental limit of digital transmission on channels with white noise is given by the classic channel capacity formula $C = W \log_2 (1 + S / N)$, where $C$ is the capacity in bit/s, $W$ is the channel bandwidth in Hz, and $S/N$ is the signal to noise ratio at the receiver.

However, finding an error correction system able to achieve this limit meant extensive research for several decades. After more than forty years of research the concept of turbo coding developed by Claude Berrou and Alain Glavieux [4] finally proved that it was possible to reach the limit of channel capacity with an encoding scheme that could be constructed in practice.

While turbo coding is not the only technique known to be able to attain the channel capacity limit [5] it is certainly the most commonly used channel coding technique for data channels in contemporary mobile communications systems.

According to the inventors the turbo coding principle was born from the experimentation with the feedback concept applied to the error correcting problem using convolutional codes [6]. At the core of a turbo coding system there is a fundamental constitutive element called interleaver. An interleaver is a system that changes the positions of input data according to an established position permutation algorithm. Inside the turbo coding process the function of the interleaving block is to help in providing codes vectors with the highest possible level of randomness (ideally, independent vectors) [7] so that the resulting code resembles as close as possible the concept of random coding used by C . E. Shannon in [8] to prove the channel capacity theorem. Therefore the interleaver is a fundamental element for the performance of a turbo code [9].

The understanding of interleaving is a subject of high interest to the specification of physical layers for both wired and wireless transmission technologies. The aim of this article is to show in detail how the turbo interleavers defined in the

cdma2000 EV-DO Revision B (3GPP2 C.S0024-B V2.0) [10] and W-CDMA (3GPP TS 25.212 V7.6.0) [11] standards work, what are their main characteristics, and what are the design principles used by each one. This article has been written with a tutorial approach in mind.

The article is organized as follows: Section 2 provides an overview of the turbo interleavers defined in cdma2000 and W-CDMA standards. In Section 3 two detailed examples of the turbo interleaving done by each standard are described. Section 4 shows a dispersion analysis for each interleaver. Finally, in section 5 the main findings and observations of this work are presented.

For a full discussion of the theory of the interleavers described in this article references [12] and [13] can be consulted.

## 2. Description of the cdma2000 and W-CDMA turbo interleavers

W-CDMA and cdma2000 use different strategies for the interleaving carried out by its turbo coding systems. The cdma2000 interleaver is based on the principle of generating the interleaving positions through a counter that generates addresses which are modified through a preset table and a function that reverses the order of the bits. The resulting address vectors determine the permutation of the input data. In cdma2000 the input to the interleaver and the output data form the interleaver are defined as arrays (vectors) of length *Nturbo*. The values that the *Nturbo* variable can take are defined by the standard.

In W-CDMA the input and output of the interleaver are treated as matrices whose dimensions (rows and columns) depend on the total length of the input data, *K*. The values the variable *K* can take are also defined by the standard. The interleaving process takes place in two steps. First the positions of the bits for each row are permuted. Then the row positions are permuted (without changing the bits in each row). In summary, the American standard (cdma2000) treats input bits as an array (or vector), while the European standard (W-CDMA) treats the input data as a matrix. To change positions the American standard uses a counter while the European one uses permutation patterns of rows and columns. The following sub-sections describe in detail the operation of the cdma2000 and W-CDMA interleavers.

### 2.1 cdma2000 turbo interleaver

Figure 1 shows the flow diagram of the interleaver used by the cdma2000 turbo encoder, which has as input the packet_size variable which is used to determine from Table 1 both then *n* y *Nturbo* parameters. The value of *n* is a interleaving parameter defined as an integer in the range $3 \le n \le 7$ . *Nturbo* is the actual number of information bits in the interleaving block and must satisfy the relationship $Nturbo \le 2^{n+5}$ .

The packet size is six bit longer than *Nturbo* because the six trail bits are used to force the turbo encoder to the initial state after the codification of the *Nturbo* data bits is complete.

**Table 1.** Turbo interleaver Parameter

| $Packet_size$ | $n$ | $N_{turbo}$ |
|---|---|---|
| 256 | 3 | 250 |
| 512 | 4 | 506 |
| 1024 | 5 | 1018 |
| 2048 | 6 | 2042 |
| 4096 | 7 | 4090 |

The parameters in Table 1 are defined for reverse link channels i.e. channels going from the mobile station to the base station. For forward link channels the (base station to mobile) $n$ is in the range $5 \leq n \leq 7$ and the values of *packet size* and *Nturbo* are different. The interleaving algorithm is the same for the reverse link as well as the direct link.

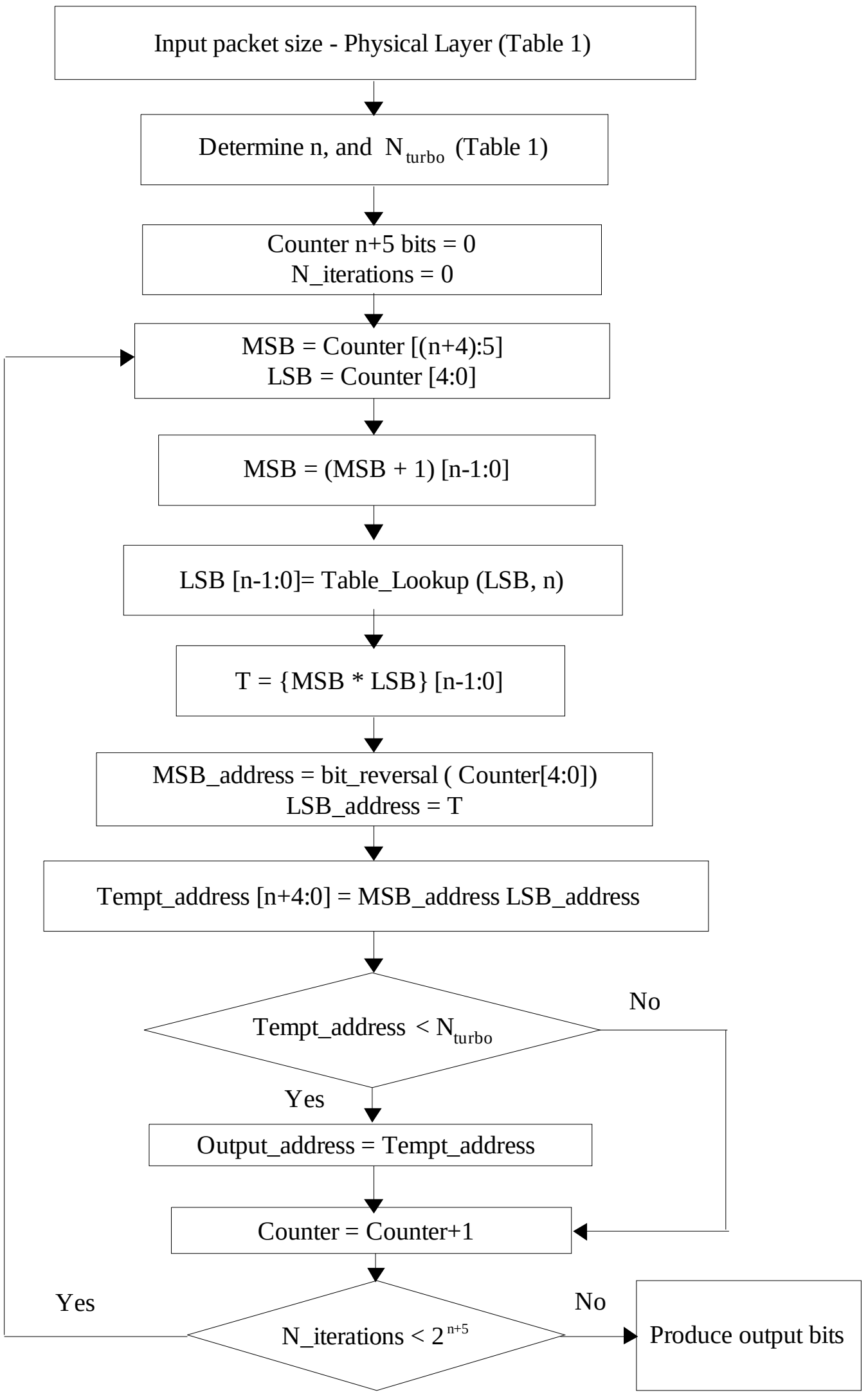

**Fig. 1.** Flow diagram for the cdma2000 standard's turbo interleaver algorithm

Figure 1 shows the sequential tasks that are performed within the interleaver proposed in the cdma2000 standard. The first task is to calculate the *MSB* address by taking the *n* least significant bits of the value of the address counter *n* most significant bits plus one.

Then Table 2 is indexed using the counter's five least significant bits. This lookup table indexing provides an *LSB* address of *n* bits.

**Table 2.** cdma2000 turbo interleaver lookup table

| *Index* | *n=3* | *n=4* | *n=5* | *n=6* | *n=7* |
|---|---|---|---|---|---|
| 0 | 1 | 5 | 27 | 3 | 15 |
| 1 | 1 | 15 | 3 | 27 | 127 |
| 2 | 3 | 5 | 1 | 15 | 89 |
| 3 | 5 | 15 | 15 | 13 | 1 |
| 4 | 1 | 1 | 13 | 29 | 31 |
| 5 | 5 | 9 | 17 | 5 | 15 |
| 6 | 1 | 9 | 23 | 1 | 61 |
| 7 | 5 | 15 | 13 | 31 | 47 |
| 8 | 3 | 13 | 9 | 3 | 127 |
| 9 | 5 | 15 | 3 | 9 | 17 |
| 10 | 3 | 7 | 15 | 15 | 119 |
| 11 | 5 | 11 | 3 | 31 | 15 |
| 12 | 3 | 15 | 13 | 17 | 57 |
| 13 | 5 | 3 | 1 | 5 | 123 |
| 14 | 5 | 15 | 13 | 39 | 95 |
| 15 | 1 | 5 | 29 | 1 | 5 |
| 16 | 3 | 13 | 21 | 19 | 85 |
| 17 | 5 | 15 | 19 | 27 | 17 |
| 18 | 3 | 9 | 1 | 15 | 55 |
| 19 | 5 | 3 | 3 | 13 | 57 |
| 20 | 3 | 1 | 29 | 45 | 15 |
| 21 | 5 | 3 | 17 | 5 | 41 |
| 22 | 5 | 15 | 25 | 33 | 93 |
| 23 | 5 | 1 | 29 | 15 | 87 |
| 24 | 1 | 13 | 9 | 13 | 63 |
| 25 | 5 | 1 | 13 | 9 | 15 |
| 26 | 1 | 9 | 23 | 15 | 13 |
| 27 | 5 | 15 | 13 | 31 | 15 |
| 28 | 3 | 11 | 13 | 17 | 81 |
| 29 | 5 | 3 | 1 | 5 | 57 |
| 30 | 5 | 15 | 13 | 15 | 31 |
| 31 | 3 | 5 | 13 | 33 | 69 |

The next step is to take the n least significant bits of the product of the previously obtained *MSB* and *LSB* addresses which will constitute the lower part, namely *LSB_address*, of the final address.

The higher part of the final address, namely *MSB_address*, is obtained by bit-reversing the five least significant bits of the counter.

*MSB_address* and *LSB_address* are then concatenated forming the final address which is stored at the *Output_address* vector if the final address is less than *Nturbo*, otherwise the address is discarded.

The counter is increased by one and the process is repeated until the *Nturbo* interleaving addresses are obtained.

The algorithm is designed so that with $2^{n+5}$ iterations is always possible to obtain the *Nturbo* required addresses, i.e. it is not possible for the iterations to end without having obtained all the interleaving addresses.

### 2.2 W-CDMA turbo interleaver

The turbo interleaver of the W-CDMA standard is based on a rectangular input bit matrix. This matrix is permuted both by columns and rows before the output bits are delivered. The input bits are denoted as $x_1, x_2, x_3, x_4, x_5, \ldots, x_K$ where $K$ is the number of input bits, where $40 \le K \le 5114$ .

Figure 2 shows a flow diagram of the algorithm used by the W-CDMA turbo interleaver.

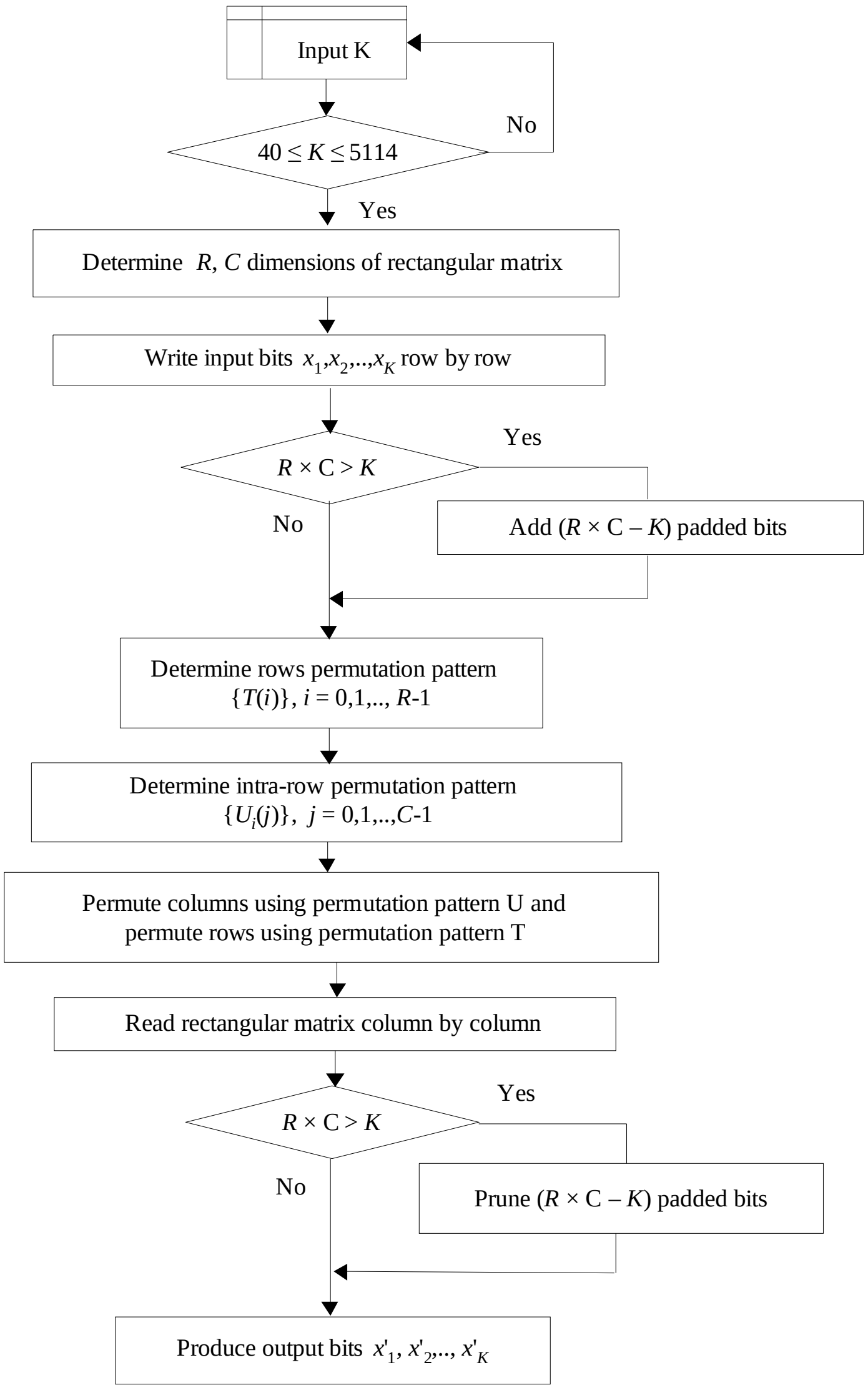

**Fig. 2.** Flow diagram for the W-CDMA standard's turbo interleaver algorithm

As shown in Fig. 2, after verifying that the length of $K$ is in the range established by the standard, then both the number of rows $R$, and the number of columns $C$ of the rectangular matrix are determined according to the rules given in Tables 3 and 4.

**Table 3.** Rules for determining the number of rows *R*

| *R* | *K* |
|---|---|
| 5 | $40 \le K \le 159$ |
| 10 | $(160 \le K \le 200) \vee (481 \le K \le 530)$ |
| 20 | *K = any other value* |

**Table 4.** Rules for determining the number of columns *C*

| *K* | *P* | *C* |
|---|---|---|
| $481 \le K \le 530$ | 53 | $C = p = 53$ |
| $K \le R \times (p+1)$ | Minimum *p* (see Table 5) | $C = p-1 \quad \text{if } (K \le R \times (p-1))$<br>$C = p \quad \text{if } (R \times (p-1) < K \le R \times p)$<br>$C = p+1 \quad \text{if } ((R \times p) < K)$ |

As shown in Table 4 when *K* is outside the range $481 \le K \le 530$, *p* is the lowest prime number such that $K \le R \times (p+1)$ and *C* is calculated according the third column of Table 4. The values for the prime number *p* are shown in Table 5.

**Table 5.** List of prime number *p* and primitive roots *v*

| *p* | *v* | *p* | *v* | *p* | *v* | *p* | *v* | *p* | *v* |
|---|---|---|---|---|---|---|---|---|---|
| 7 | 3 | 47 | 5 | 101 | 2 | 157 | 5 | 223 | 3 |
| 11 | 2 | 53 | 2 | 103 | 5 | 163 | 2 | 227 | 2 |
| 13 | 2 | 59 | 2 | 107 | 2 | 167 | 5 | 229 | 6 |
| 17 | 3 | 61 | 2 | 109 | 6 | 173 | 2 | 233 | 3 |
| 19 | 2 | 67 | 2 | 113 | 3 | 179 | 2 | 239 | 7 |
| 23 | 5 | 71 | 7 | 127 | 3 | 181 | 2 | 241 | 7 |
| 29 | 2 | 73 | 5 | 131 | 2 | 191 | 19 | 251 | 6 |
| 31 | 3 | 79 | 3 | 137 | 3 | 193 | 5 | 257 | 3 |
| 37 | 2 | 81 | 2 | 139 | 2 | 197 | 2 | | |
| 41 | 6 | 89 | 3 | 149 | 2 | 199 | 3 | | |
| 43 | 3 | 97 | 5 | 151 | 6 | 211 | 2 | | |

The matrix is filled with the *K* input bits by rows from top to bottom. If $R \times C > K$ , then the matrix is zero (or one) padded.

The row permutation pattern which is represented by the vector $\langle T(i) \rangle_{i \in \{0,1,\ldots,R-1\}}$ and depends on the number of input bits *K* as showed in Table 6.

**Table 6.** Rules for determining the row permutation pattern

| *K* | *R* | *Inter-row permutation patterns* $\langle T(i)\rangle_{i\in\{0,1,\ldots,R-1\}}$ |
|---|---|---|
| $40 \le K \le 159$ | 5 | $\langle 4,3,2,1,0\rangle$ |
| $(160 \le K \le 200) \vee (481 \le K \le 530)$ | 10 | $\langle 9,8,7,6,5,4,3,2,1,0\rangle$ |
| $(2281 \le K \le 2480) \vee (3161 \le K \le 3210)$ | 20 | $\langle 19,9,14,4,0,2,5,7,12,18,16,13,17,15,3,1,6,11,8,10\rangle$ |
| $K = \text{any other value}$ | 20 | $\langle 19,9,14,4,0,2,5,7,12,18,10,8,13,17,3,1,16,6,15,11\rangle$ |

The next task is to calculate the permutation pattern for the columns in each row. This pattern is defined by the matrix $\langle U_i(j)\rangle_{j\in\{0,1,\ldots,C-1\}}$ as indicated in Table 7.

**Table 7.** Column permutations Patterns

| *Condition* | *Column permutation pattern* $\langle U_i(j)\rangle_{j\in\{0,1,\ldots,C-1\}}$ |
|---|---|
| *C = p* | $U_i(j) = s((j\times r_i)\mathrm{mod}(p-1)),\ j = 0,1,\ldots(p-2), U_i(p-1) = 0$ |
| *C = p+1* | $U_i(j) = s((j\times r_i)\mathrm{mod}(p-2)),\ j = 0,1,\ldots(p-2), U_i(p-1) = 0\ and\ U_i(p) = p$ <br> $K = R\times C$, exchange $U_{R-1}(p)$ with $U_{R-1}(0)$ |
| *C = p-1* | $U_i(j) = s((j\times r_i)\mathrm{mod}(p-1)),\ j = 0,1,\ldots(p-2)$ |

In Table 7 the variable *s* corresponds to the base sequence $\langle s(j)\rangle_{j\in\{0,1,\ldots,p-2\}}$ used to generate the column permutation pattern and is defined by the eq. (1):

$$s\ (j) = s((v\times s(j-1)\ )\mathrm{mod}(p)),\ j = 1,2,\ldots(p-2), s(0) = 1 \quad (1)$$

where *v* is the primitive root as defined in Table 5.

The variable $r_i$ is the sequence of permuted prime integers defined in equation (2):

$$r_{T(i)} = q_i, i = 0,1,\ldots, R-1 \quad (2)$$

Where the subscript $T_{(i)}$ is the row permutation vector given in Table 6 and *q* is the sequence of primes determined by the lowest primes $q_i$ such that $g.c.d\ (q_i, p-1) = 1, q_i > 6$ and, $q_i > q_{(i-1)}, i = 1,2,\ldots, R-1.$

The original input matrix is first column permuted with the column permutation pattern $\langle U_i(j)\rangle_{j\in\{0,1,\ldots,C-1\}}$ and then with the row permutation patterns $\langle T(i)\rangle_{i\in\{0,1,\ldots,R-1\}}$. Finally data are read by columns from left to right. If extra zero (or one) bits where initially padded these bits must be removed (the number of bits removed equals $(R \times C) - K$, producing the bits $x'_1, x'_2, x'_3, x'_4, x'_5, \ldots, x'_K$ .

## 3. Examples of operation

### 3.1 Turbo interleaver standard cdma2000

In this example a packet size $packet_size = 256$ is used. Following the algorithm described in Fig. 1 we have:

1. From Table 1 *n* =3 and *Nturbo* = 250.
2. Start the counter of *n*+5=8 bits at zero, i.e. counter = 00000000.
3. Save the three most significant bits of counter in the *MSB* variable.
4. Save the five least significant bits of counter in the *LSB* variable.
5. Add 1 to *MSB*, i.e. *MSB* = 001 for the first iteration
6. Use Table 2 with the five *LSBs* of counter and column *n* (*index* = 00000 for the first iteration) and store the *n*-bit result in *LSB*, i.e. *LSB* = 001 for the first iteration.
7. Take the *n* least significant bits of the *MSB* x *LSB* product and store it in *T*.
8. Bit reverse the five least significant bits of counter and save it in *MSB_address*, i.e. *MSB_address* = 00000 for the first iteration
9. Concatenate *MSB_address* and *LSB_address*=*T* into *Temp_address*, i.e. *Temp_address* =00000001 for the first iteration
10. If *Temp_address* < *Nturbo* then deliver valid *output_address*, i.e. *output_address*= 00000001 for the first iteration
11. Add 1 to counter
12. Go to step 3

Table 8 shows the six addresses which are discarded during the process because of being greater than the value of *Nturbo*.

**Table 8.** Example of invalid interleaving addresses

| *Address Counter MSB LSB* | *MSB = MSB+1* | *LSB = Table2[LSB,n]* | *MSB_address = reverse(LSB counter)* | *LSB_address = MSB x LSB* | *Tempt_address = [MSB][LSB]* |
|---|---|---|---|---|---|
| 000 11111 | 001 | 011 | 11111 | 001x011 = 011 | 11111011 = 251base 10 |
| 001 11111 | 010 | 011 | 11111 | 010x011 = 110 | 11111110 = 254base 10 |

| | | | | | | |
|---|---|---|---|---|---|---|
| 011 | 11111 | 100 | 011 | 11111 | 100x011 = 100 | 11111100 = 252base10 |
| 101 | 11111 | 110 | 011 | 11111 | 110x011 = 010 | 11111010 = 250base10 |
| 110 | 11111 | 111 | 011 | 11111 | 111x011 = 101 | 11111101 = 253base10 |
| 100 | 11111 | 101 | 011 | 11111 | 101x011 = 111 | 11111111 = 255base10 |

For this example, the input vector has 250 data, whose values for simplicity are consecutively numbered from one to two hundred and fifty. As the input vector is relatively large in size only some positions with their values are shown in Table 9. Figure 3 shows the interleaved output data vector.

**Table 9.** cdma2000 example input vector

| | | | | | | | | | | |
|---|---|---|---|---|---|---|---|---|---|---|
| 1 | 2 | 3 | … | 16 | ... | 242 | … | 248 | 249 | 250 |

| | | | | | | | | | | | | | | | |
|---|---|---|---|---|---|---|---|---|---|---|---|---|---|---|---|
| **{**2 | 130 | 68 | 198 | 34 | 166 | 98 | 230 | 20 | 150 | 84 | 214 | 52 | 182 | 118 | 242 |
| 12 | 142 | 76 | 206 | 44 | 174 | 110 | 238 | 26 | 158 | 90 | 222 | 60 | 190 | 126 | 3 |
| 131 | 71 | 195 | 35 | 163 | 99 | 227 | 23 | 147 | 87 | 211 | 55 | 179 | 115 | 243 | 15 |
| 139 | 79 | 203 | 47 | 171 | 107 | 235 | 27 | 155 | 91 | 219 | 63 | 187 | 123 | 4 | 132 |
| 66 | 200 | 36 | 168 | 100 | 232 | 18 | 152 | 82 | 216 | 50 | 184 | 120 | 244 | 10 | 144 |
| 74 | 208 | 42 | 176 | 112 | 240 | 28 | 160 | 92 | 224 | 58 | 192 | 128 | 250 | 5 | 133 |
| 69 | 197 | 37 | 165 | 101 | 229 | 21 | 149 | 85 | 213 | 53 | 181 | 117 | 245 | 13 | 141 |
| 77 | 205 | 45 | 173 | 109 | 237 | 29 | 157 | 93 | 221 | 61 | 189 | 125 | 6 | 134 | 72 |
| 194 | 38 | 162 | 102 | 226 | 24 | 146 | 88 | 210 | 56 | 178 | 114 | 246 | 16 | 138 | 80 |
| 202 | 48 | 170 | 106 | 234 | 30 | 154 | 94 | 218 | 64 | 186 | 122 | 7 | 135 | 67 | 199 |
| 39 | 167 | 103 | 231 | 19 | 151 | 83 | 215 | 51 | 183 | 119 | 247 | 11 | 143 | 75 | 207 |
| 43 | 175 | 111 | 239 | 31 | 159 | 95 | 223 | 59 | 191 | 127 | 8 | 136 | 70 | 196 | 40 |
| 164 | 104 | 228 | 22 | 148 | 86 | 212 | 54 | 180 | 116 | 248 | 14 | 140 | 78 | 204 | 46 |
| 172 | 108 | 236 | 32 | 156 | 96 | 220 | 62 | 188 | 124 | 1 | 129 | 65 | 193 | 33 | 161 |
| 97 | 225 | 17 | 145 | 81 | 209 | 49 | 177 | 113 | 241 | 9 | 137 | 73 | 201 | 41 | 169 |
| 105 | 233 | 25 | 153 | 89 | 217 | 57 | 185 | 121 | 249**}** | | | | | | |

**Fig. 3.** cdma2000 output data after interleaving

As can be seen in Figure 3 the input data has been totally interleaved from their original positions. For instance at position 16 is the element 242 and at position 242 is the element 233, whereas at the same positions in the input vector in Table 9 are the data 16 and 242 respectively.

Figure 4 shows the results of this example plotted in a Cartesian plane, where the x-axis represents the index and the y-axis the position of the output data.

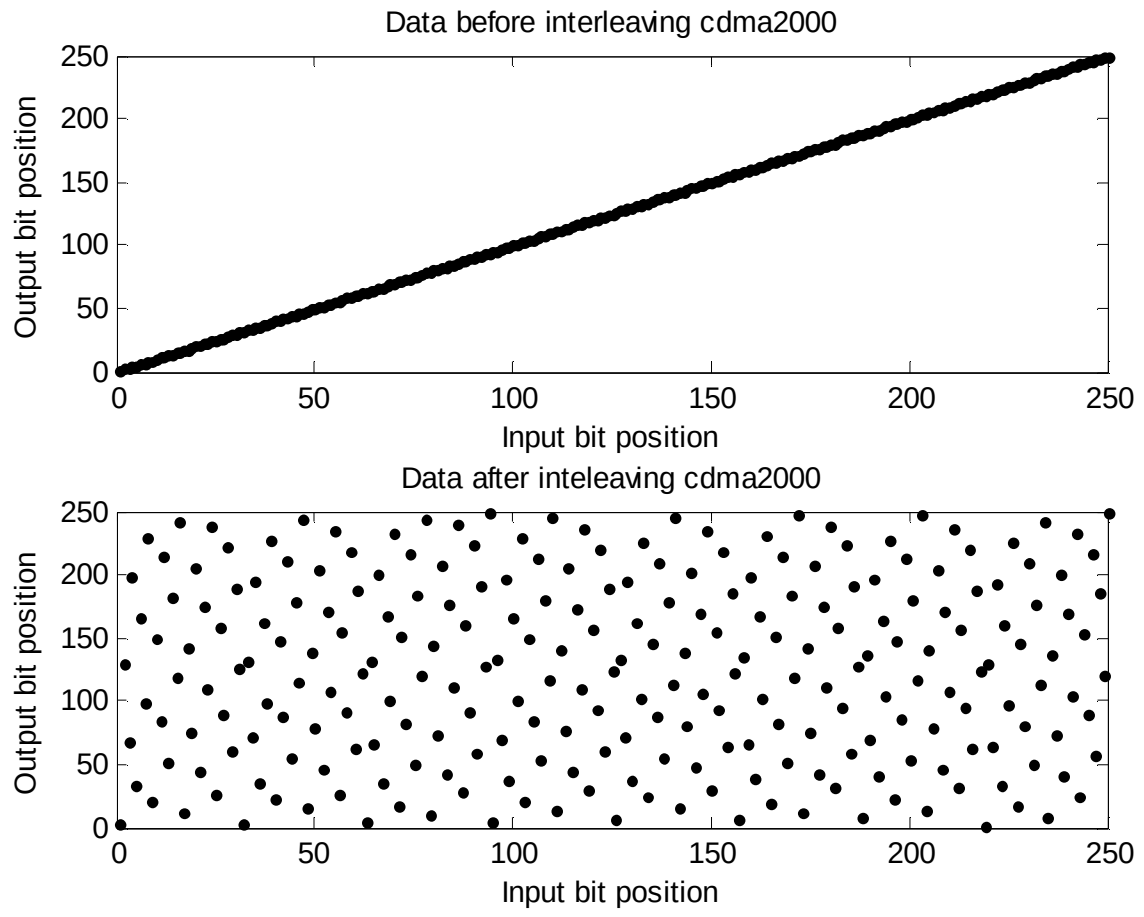

Fig. 4. Input vector vs output vector before and after interleaving

**3.2 Turbo interleaver standard W-CDMA**

For this example a value of $K = 250$ input data is used. From Table 3 the number of rows *R* must be 20 and from Table 4 the number of columns *C* must be 13. This value for *C* is found as follows: from Table 5 the smallest prime such that $250 \le 20 \times (p+1)$ is *p* = 13; according to the rules in Table 4 as $20 \times 12 < 250 \le 20 \times 13$ meets the condition $R \times (p+1) < K \le R \times p$, then $C = p = 13$. The rectangular matrix has then $20 \times 13$ = 260 elements. In this example the input values, for simplicity are chosen to be the integers from 1 to 250 which are written by rows into the input matrix. Since the matrix size is greater than *K* the empty positions are filled with zeros as shown in Table 10.

According to the rules in Table 6 the row permutation pattern $\langle T(i) \rangle_{i \in \{0,1,\ldots,19\}}$ for this case is $T = \langle 19,9,14,4,0,2,5,7,12,18,10,8,13,17,3,1,16,6,15,11 \rangle$. Similarly, as in this example $C = p$, the column permutation patterns according to the rules in Table 7 are determined by equation (3):

$$U_i(j) = s((j \times r_i) \bmod (13-1)), \quad j = 0,1,\ldots,12, U_i(12) = 0 \quad (3)$$

Table 11 shows the base sequence which has been generated using Eq. (1) and which is used in Eq. (3).

To obtain the sequence of permuted prime integers $r_i$ the sequence of prime numbers *q* is needed. The sequence *q* must satisfy the condition $g.c.d\ (q_i, 13-1) = 1$, $q_i > 6$ and $q_i > q_{(i-1)}, i = 1,2,\ldots,19$, in this example giving as a result

$$q = [1,7,11,13,17,19,23,29,31,37,41,43,47,53,59,61,67,71,73,79] \ .$$

The permuted sequence $r_i$ is then:

$$r_{T_{(i)}} = q_i, i = 0,1,...,19, r_{T_{(i)}} = [17,61,19,59,13,23,71,29,43,7,41,79,31,47,11,73,67,53,37,1]$$

Table 12 shows the column permutation patterns produced by Eq. (3).

Table 13 shows the rectangular matrix after applying the permutation pattern *U* and Table 14 after applying the permutation pattern *T*. Table 15 shows the resulting output after reading the matrix column by column. It can be observed at the first column of Table 15 the six fill-in zeros. Figure 5 shows the end result after pruning the fill-in zeros. Figure 6 shows this example's results plotted in a Cartesian plane where the x-axis represents the index and the y-axis the position of the output data.

**Table 10.** 20x13 matrix before interleaving

| | | | | | | | | | | | | |
|---|---|---|---|---|---|---|---|---|---|---|---|---|
| 1 | 2 | 3 | 4 | 5 | 6 | 7 | 8 | 9 | 10 | 11 | 12 | 13 |
| 14 | 15 | 16 | 17 | 18 | 19 | 20 | 21 | 22 | 23 | 24 | 25 | 26 |
| 27 | 28 | 29 | 30 | 31 | 32 | 33 | 34 | 35 | 36 | 37 | 38 | 39 |
| 40 | 41 | 42 | 43 | 44 | 45 | 46 | 47 | 48 | 49 | 50 | 51 | 52 |
| 53 | 54 | 55 | 56 | 57 | 58 | 59 | 60 | 61 | 62 | 63 | 64 | 65 |
| 66 | 67 | 68 | 69 | 70 | 71 | 72 | 73 | 74 | 75 | 76 | 77 | 78 |
| 79 | 80 | 81 | 82 | 83 | 84 | 85 | 86 | 87 | 88 | 89 | 90 | 91 |
| 92 | 93 | 94 | 95 | 96 | 97 | 98 | 99 | 100 | 101 | 102 | 103 | 104 |
| 105 | 106 | 107 | 108 | 109 | 110 | 111 | 112 | 113 | 114 | 115 | 116 | 117 |
| 118 | 119 | 120 | 121 | 122 | 123 | 124 | 125 | 126 | 127 | 128 | 129 | 130 |
| 131 | 132 | 133 | 134 | 135 | 136 | 137 | 138 | 139 | 140 | 141 | 142 | 143 |
| 144 | 145 | 146 | 147 | 148 | 149 | 150 | 151 | 152 | 153 | 154 | 155 | 156 |
| 157 | 158 | 159 | 160 | 161 | 162 | 163 | 164 | 165 | 166 | 167 | 168 | 169 |
| 170 | 171 | 172 | 173 | 174 | 175 | 176 | 177 | 178 | 179 | 180 | 181 | 182 |
| 183 | 184 | 185 | 186 | 187 | 188 | 189 | 190 | 191 | 192 | 193 | 194 | 195 |
| 196 | 197 | 198 | 199 | 200 | 201 | 201 | 203 | 204 | 205 | 206 | 207 | 208 |
| 209 | 210 | 211 | 212 | 213 | 214 | 215 | 216 | 217 | 218 | 219 | 220 | 221 |
| 222 | 223 | 224 | 225 | 226 | 227 | 228 | 229 | 230 | 231 | 232 | 233 | 234 |
| 235 | 236 | 237 | 238 | 239 | 240 | 241 | 242 | 243 | 244 | 245 | 246 | 247 |
| 248 | 249 | 250 | 0 | 0 | 0 | 0 | 0 | 0 | 0 | 0 | 0 | 0 |

**Table 11.** Base sequence for row permutations $\langle s(j) \rangle_{j \in \{0,1,...,11\}}$

| | | | | | | | | | | | |
|---|---|---|---|---|---|---|---|---|---|---|---|
| 1 | 2 | 4 | 8 | 3 | 6 | 12 | 11 | 9 | 5 | 10 | 7 |

**Table 12.** Column permutation matrix $U_i(j)$

| | | | | | | | | | | | | |
|---|---|---|---|---|---|---|---|---|---|---|---|---|
| 2 | 7 | 11 | 9 | 10 | 3 | 13 | 8 | 4 | 6 | 5 | 12 | 1 |
| 2 | 3 | 5 | 9 | 4 | 7 | 13 | 12 | 10 | 6 | 11 | 8 | 1 |
| 2 | 12 | 5 | 6 | 4 | 8 | 13 | 3 | 10 | 9 | 11 | 7 | 1 |
| 2 | 8 | 11 | 6 | 10 | 12 | 13 | 7 | 4 | 9 | 5 | 3 | 1 |
| 2 | 3 | 5 | 9 | 4 | 7 | 13 | 12 | 10 | 6 | 11 | 8 | 1 |
| 2 | 8 | 11 | 6 | 10 | 12 | 13 | 7 | 4 | 9 | 5 | 3 | 1 |
| 2 | 8 | 11 | 6 | 10 | 12 | 13 | 7 | 4 | 9 | 5 | 3 | 1 |
| 2 | 7 | 11 | 9 | 10 | 3 | 13 | 8 | 4 | 6 | 5 | 12 | 1 |
| 2 | 12 | 5 | 6 | 4 | 8 | 13 | 3 | 10 | 9 | 11 | 7 | 1 |
| 2 | 12 | 5 | 6 | 4 | 8 | 13 | 3 | 10 | 9 | 11 | 7 | 1 |
| 2 | 7 | 11 | 9 | 10 | 3 | 13 | 8 | 4 | 6 | 5 | 12 | 1 |
| 2 | 12 | 5 | 6 | 4 | 8 | 13 | 3 | 10 | 9 | 11 | 7 | 1 |
| 2 | 12 | 5 | 6 | 4 | 8 | 13 | 3 | 10 | 9 | 11 | 7 | 1 |
| 2 | 8 | 11 | 6 | 10 | 12 | 13 | 7 | 4 | 9 | 5 | 3 | 1 |
| 2 | 8 | 11 | 6 | 10 | 12 | 13 | 7 | 4 | 9 | 5 | 3 | 1 |
| 2 | 3 | 5 | 9 | 4 | 7 | 13 | 12 | 10 | 6 | 11 | 8 | 1 |
| 2 | 12 | 5 | 6 | 4 | 8 | 13 | 3 | 10 | 9 | 11 | 7 | 1 |
| 2 | 7 | 11 | 9 | 10 | 3 | 13 | 8 | 4 | 6 | 5 | 12 | 1 |
| 2 | 3 | 5 | 9 | 4 | 7 | 13 | 12 | 10 | 6 | 11 | 8 | 1 |
| 2 | 3 | 5 | 9 | 4 | 7 | 13 | 12 | 10 | 6 | 11 | 8 | 1 |

**Table 13.** Rectangular matrix after column interleaving.

| | | | | | | | | | | | | |
|---|---|---|---|---|---|---|---|---|---|---|---|---|
| 2 | 7 | 11 | 9 | 10 | 3 | 13 | 8 | 4 | 6 | 5 | 12 | 1 |
| 15 | 16 | 18 | 22 | 17 | 20 | 26 | 25 | 23 | 19 | 24 | 21 | 14 |
| 28 | 38 | 31 | 32 | 30 | 34 | 39 | 29 | 36 | 35 | 37 | 33 | 27 |
| 41 | 47 | 50 | 45 | 49 | 51 | 52 | 46 | 43 | 48 | 44 | 42 | 40 |
| 54 | 55 | 57 | 61 | 56 | 59 | 65 | 64 | 62 | 58 | 63 | 60 | 53 |
| 67 | 73 | 76 | 71 | 75 | 77 | 78 | 72 | 69 | 74 | 70 | 68 | 66 |
| 80 | 86 | 89 | 84 | 88 | 90 | 91 | 85 | 82 | 87 | 83 | 81 | 79 |
| 93 | 98 | 102 | 100 | 101 | 94 | 104 | 99 | 95 | 97 | 96 | 103 | 92 |
| 106 | 116 | 109 | 110 | 108 | 112 | 117 | 107 | 114 | 113 | 115 | 111 | 105 |
| 119 | 129 | 122 | 123 | 121 | 125 | 130 | 120 | 127 | 126 | 128 | 124 | 118 |
| 132 | 137 | 141 | 139 | 140 | 133 | 143 | 138 | 134 | 136 | 135 | 142 | 131 |
| 145 | 155 | 148 | 149 | 147 | 151 | 156 | 146 | 153 | 152 | 154 | 150 | 144 |
| 158 | 168 | 161 | 162 | 160 | 164 | 169 | 159 | 166 | 165 | 167 | 163 | 157 |
| 171 | 177 | 180 | 175 | 179 | 181 | 182 | 176 | 173 | 178 | 174 | 172 | 170 |
| 184 | 190 | 193 | 188 | 192 | 194 | 195 | 189 | 186 | 191 | 187 | 185 | 183 |
| 197 | 198 | 200 | 204 | 199 | 202 | 208 | 207 | 205 | 201 | 206 | 203 | 196 |
| 210 | 220 | 213 | 214 | 212 | 216 | 221 | 211 | 218 | 217 | 219 | 215 | 209 |
| 223 | 228 | 232 | 230 | 231 | 224 | 234 | 229 | 225 | 227 | 226 | 233 | 222 |
| 236 | 237 | 239 | 243 | 238 | 241 | 247 | 246 | 244 | 240 | 245 | 242 | 235 |
| 249 | 250 | 0 | 0 | 0 | 0 | 0 | 0 | 0 | 0 | 0 | 0 | 248 |

**Table 14** Rectangular matrix after both column and row interleaving

| | | | | | | | | | | | | |
|---|---|---|---|---|---|---|---|---|---|---|---|---|
| 249 | 250 | 0 | 0 | 0 | 0 | 0 | 0 | 0 | 0 | 0 | 0 | 248 |
| 119 | 129 | 122 | 123 | 121 | 125 | 130 | 120 | 127 | 126 | 128 | 124 | 118 |
| 184 | 190 | 193 | 188 | 192 | 194 | 195 | 189 | 186 | 191 | 187 | 185 | 183 |
| 54 | 55 | 57 | 61 | 56 | 59 | 65 | 64 | 62 | 58 | 63 | 60 | 53 |
| 2 | 7 | 11 | 9 | 10 | 3 | 13 | 8 | 4 | 6 | 5 | 12 | 1 |
| 28 | 38 | 31 | 32 | 30 | 34 | 39 | 29 | 36 | 35 | 37 | 33 | 27 |
| 67 | 73 | 76 | 71 | 75 | 77 | 78 | 72 | 69 | 74 | 70 | 68 | 66 |
| 93 | 98 | 102 | 100 | 101 | 94 | 104 | 99 | 95 | 97 | 96 | 103 | 92 |
| 158 | 168 | 161 | 162 | 160 | 164 | 169 | 159 | 166 | 165 | 167 | 163 | 157 |
| 236 | 237 | 239 | 243 | 238 | 241 | 247 | 246 | 244 | 240 | 245 | 242 | 235 |
| 132 | 137 | 141 | 139 | 140 | 133 | 143 | 138 | 134 | 136 | 135 | 142 | 131 |
| 106 | 116 | 109 | 110 | 108 | 112 | 117 | 107 | 114 | 113 | 115 | 111 | 105 |
| 171 | 177 | 180 | 175 | 179 | 181 | 182 | 176 | 173 | 178 | 174 | 172 | 170 |
| 223 | 228 | 232 | 230 | 231 | 224 | 234 | 229 | 225 | 227 | 226 | 233 | 222 |
| 41 | 47 | 50 | 45 | 49 | 51 | 52 | 46 | 43 | 48 | 44 | 42 | 40 |
| 15 | 16 | 18 | 22 | 17 | 20 | 26 | 25 | 23 | 19 | 24 | 21 | 14 |
| 210 | 220 | 213 | 214 | 212 | 216 | 221 | 211 | 218 | 217 | 219 | 215 | 209 |
| 80 | 86 | 89 | 84 | 88 | 90 | 91 | 85 | 82 | 87 | 83 | 81 | 79 |
| 197 | 198 | 200 | 204 | 199 | 202 | 208 | 207 | 205 | 201 | 206 | 203 | 196 |
| 145 | 155 | 148 | 149 | 147 | 151 | 156 | 146 | 153 | 152 | 154 | 150 | 144 |

**Table 15.** Rectangular matrix after reading column by column

| | | | | | | | | | | | |
|---|---|---|---|---|---|---|---|---|---|---|---|
| 249 | 119 | 184 | 54 | .. | .. | .. | 15 | 210 | 80 | 197 | 145 |
| 250 | 129 | 190 | 55 | .. | .. | .. | 16 | 220 | 86 | 198 | 155 |
| 0 | 122 | 193 | 57 | .. | .. | .. | 18 | 213 | 89 | 200 | 148 |
| 0 | 123 | 188 | 61 | .. | .. | .. | 22 | 214 | 84 | 204 | 149 |
| 0 | 121 | 192 | 56 | .. | .. | .. | 17 | 212 | 88 | 199 | 147 |
| 0 | 125 | 94 | 59 | .. | .. | .. | 20 | 216 | 90 | 202 | 151 |
| 0 | 130 | 195 | 65 | .. | .. | .. | 26 | 221 | 91 | 208 | 156 |
| 0 | 120 | 189 | 64 | .. | .. | .. | 25 | 211 | 85 | 207 | 146 |
| 0 | 127 | 186 | 62 | .. | .. | .. | 23 | 218 | 82 | 205 | 153 |
| 0 | 126 | 191 | 58 | .. | .. | .. | 19 | 217 | 87 | 201 | 152 |
| 0 | 128 | 187 | 63 | .. | .. | .. | 24 | 219 | 83 | 206 | 154 |
| 0 | 124 | 185 | 60 | .. | .. | .. | 21 | 215 | 81 | 203 | 150 |
| 248 | 118 | 183 | 53 | .. | .. | .. | 14 | 209 | 79 | 196 | 144 |

| | | | | | | | | | | | | | | |
|---|---|---|---|---|---|---|---|---|---|---|---|---|---|---|
| **{**249 | 119 | 184 | 54 | 2 | 28 | 67 | 93 | 158 | 236 | 132 | 106 | 171 | 223 | 41 |
| 15 | 210 | 80 | 197 | 145 | 250 | 129 | 190 | 55 | 7 | 38 | 73 | 98 | 168 | 237 |
| 137 | 116 | 177 | 228 | 47 | 16 | 220 | 86 | 198 | 155 | 122 | 193 | 57 | 11 | 31 |
| 76 | 102 | 161 | 239 | 141 | 109 | 180 | 232 | 50 | 18 | 213 | 89 | 200 | 148 | 123 |
| 188 | 61 | 9 | 32 | 71 | 100 | 162 | 243 | 139 | 110 | 175 | 230 | 45 | 22 | 214 |
| 84 | 204 | 149 | 121 | 192 | 56 | 10 | 30 | 75 | 101 | 160 | 138 | 140 | 108 | 179 |
| 231 | 49 | 17 | 212 | 88 | 199 | 147 | 125 | 194 | 59 | 3 | 34 | 77 | 94 | 164 |
| 241 | 133 | 112 | 181 | 224 | 51 | 20 | 216 | 90 | 202 | 151 | 130 | 195 | 65 | 13 |
| 39 | 78 | 104 | 169 | 247 | 143 | 117 | 182 | 234 | 52 | 26 | 221 | 91 | 208 | 156 |
| 120 | 189 | 64 | 8 | 29 | 72 | 99 | 159 | 246 | 138 | 107 | 176 | 229 | 46 | 25 |
| 211 | 85 | 207 | 146 | 127 | 186 | 62 | 4 | 36 | 69 | 95 | 166 | 244 | 134 | 114 |
| 173 | 225 | 43 | 23 | 218 | 82 | 205 | 153 | 126 | 191 | 58 | 6 | 35 | 74 | 97 |
| 165 | 240 | 136 | 113 | 178 | 227 | 48 | 19 | 217 | 87 | 201 | 152 | 128 | 187 | 63 |
| 5 | 37 | 70 | 96 | 167 | 245 | 135 | 115 | 174 | 226 | 44 | 24 | 219 | 83 | 206 |
| 154 | 124 | 185 | 60 | 12 | 33 | 68 | 103 | 163 | 242 | 142 | 111 | 172 | 233 | 42 |
| 21 | 215 | 81 | 203 | 150 | 248 | 118 | 183 | 53 | 1 | 27 | 66 | 92 | 157 | 235 |
| 131 | 105 | 170 | 222 | 40 | 14 | 209 | 79 | 196 | 144**}** | | | | | |

**Fig. 5.** W-CDMA interleaved output data

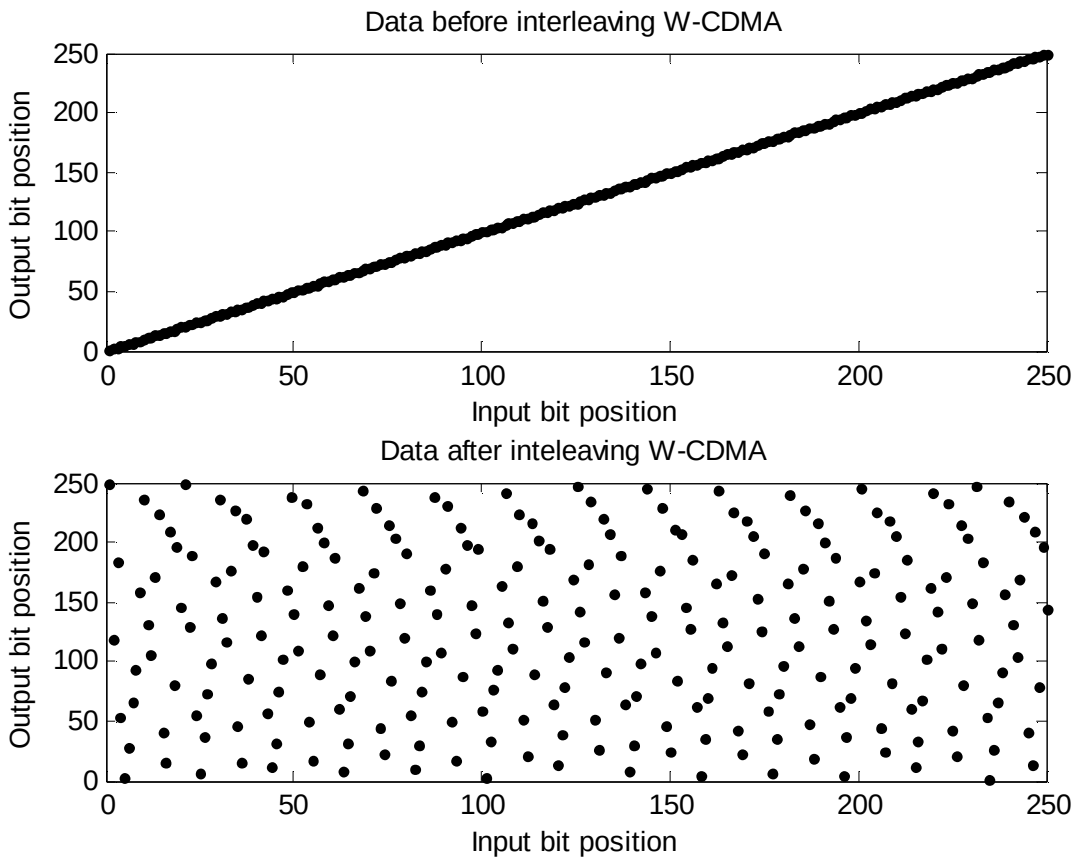

**Fig. 6.** Input vector vs output vector before and after interleaving, packet size = 250

### 3.3 Some test vectors with binary digits

As example vectors Tables 16 and 17 shows the results for the one hundred ones followed one hundred and fifty zeros vector, and the complementary vector (one hundred fifty zeros followed by one hundred ones) for W-CDMA and

cdma2000 respectively.

The notation used for both input and output vectors uses binary and hexadecimal format to avoid the ambiguity that may cause the size of the input data not being a multiple of four.

**Table 16.** W-CDMA and cdma2000 interleaving test vectors. Input data: 100 ones and 150 zeros

| Input / Output | |
|---|---|
| | $00_2$0000000000000000000000000000000000000FFFFFFFFFFFFFFFFFFFFFFFFF$_{16}$ |
| W-CDMA | $00_2$B03E0B01E1607C2C0F8581F0B03E1603C2C0F8580F0B03E1603C2C0F82C0F8$_{16}$ |
| cdma2000 | $00_2$54545554A8A8A8A951515152A2A2A2A545454545454545554A8A8AAA95151555$_{16}$ |

**Table 17.** W-CDMA and cdma2000 interleaving test vectors. Input data: 150 zeros and 100 ones

| Input / Output | |
|---|---|
| | 3FFFFFFFFFFFFFFFFFFFFFFFFC0000000000000000000000000000000000000$_{16}$ |
| W-CDMA | $01_2$4CC154CC169982D3305A66094CC169982D3304A66094CC129982D330553305$_{16}$ |
| cdma2000 | 2AAA2A2A155454542AA8AAA855515150AAA2A2A2AAA2AAA155454542AA8A8A8$_{16}$ |

In Table 16 and 17 it can easily be seen that the input data are interleaved in a different way by the turbo interleaver of each standard. For instance, in W-DCMA it appears several times consecutive ones at the output (nibbles B16 and C16, for example) but they never come from consecutive positions at the input vector. In cdma2000 it never appears for this example consecutive ones at the output vector.

**4. Dispersion analysis**

Although the output produced by interleavers (Figures 3 and 5) seem to have elements of randomness, it should be noticed that the transformation $T[x] = y$ between input and output positions is both deterministic and bijective, i.e. the mapping is fixed, one to one and for all y there is a single x and vice versa.

Because of this reason it is more useful to calculate the average interleaving distance, defined as the average value of the distance between the input and output positions, $L_{avg}$, as well the standard deviation of these distances.

Table 18 shows the values of $L_{avg}$ and standard deviation for different lengths of input data. As shown in Table 18 the values of $L_{avg}$ and standard deviation are very similar for the two compared standards.

Similarly, as noted at the histograms of interleaving distance versus frequency of Figure 7 and 8 for 250 data, the dispersion patterns are statistically similar. Patterns obtained for higher input values have similar behavior.

**Tabla 18.** $L_{avg}$ distance of interleaving

| Length of input data | $L_{avg}$ | | Standard deviation | |
|---|---|---|---|---|
| | cdma2000 | W-CDMA | cdma2000 | W-CDMA |
| 250 | 82.19 | 81.54 | 58.36 | 59.23 |
| 506 | 168.96 | 170.78 | 118.63 | 121.65 |
| 1018 | 334.60 | 333.27 | 240.71 | 240.58 |
| 2042 | 678.2 | 679.68 | 481.04 | 483.78 |
| 4090 | 1369.9 | 1358.8 | 965.65 | 964.15 |

The interleaving process does not mean that all positions must change at the output. For example, in cdma2000 for an input size equal to 506 positions 68, 84 and 338 appear at the same positions at the interleaver's output. For W-CDMA for the same input size (506) it happens the same with position 385.

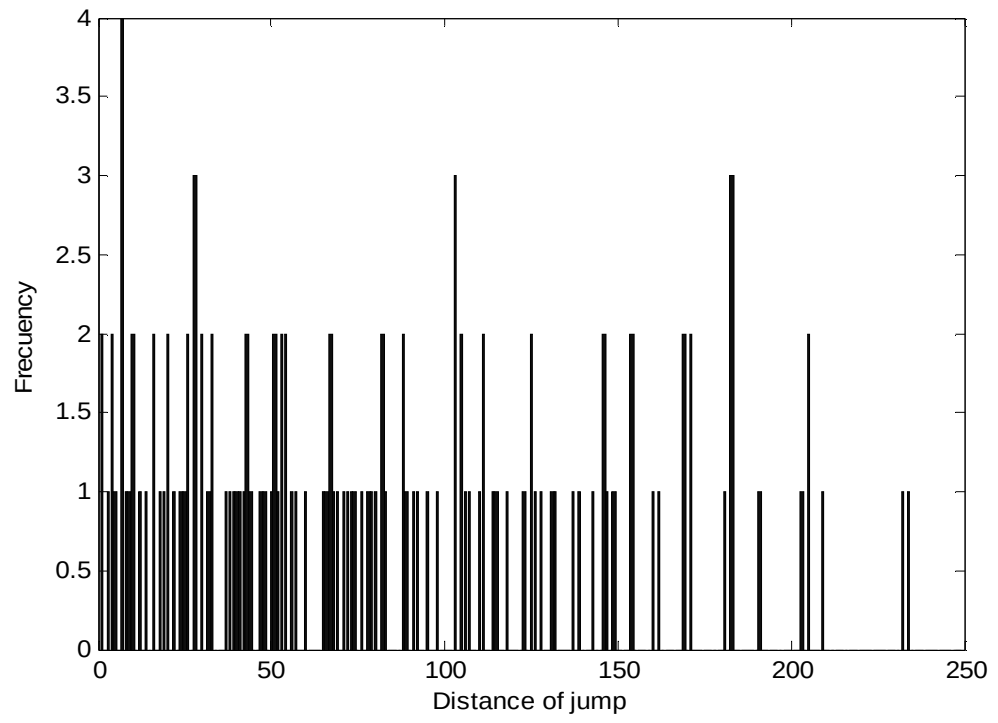

**Fig. 7.** Interleaving distance versus frequency for W-CDMA, packet size = 250

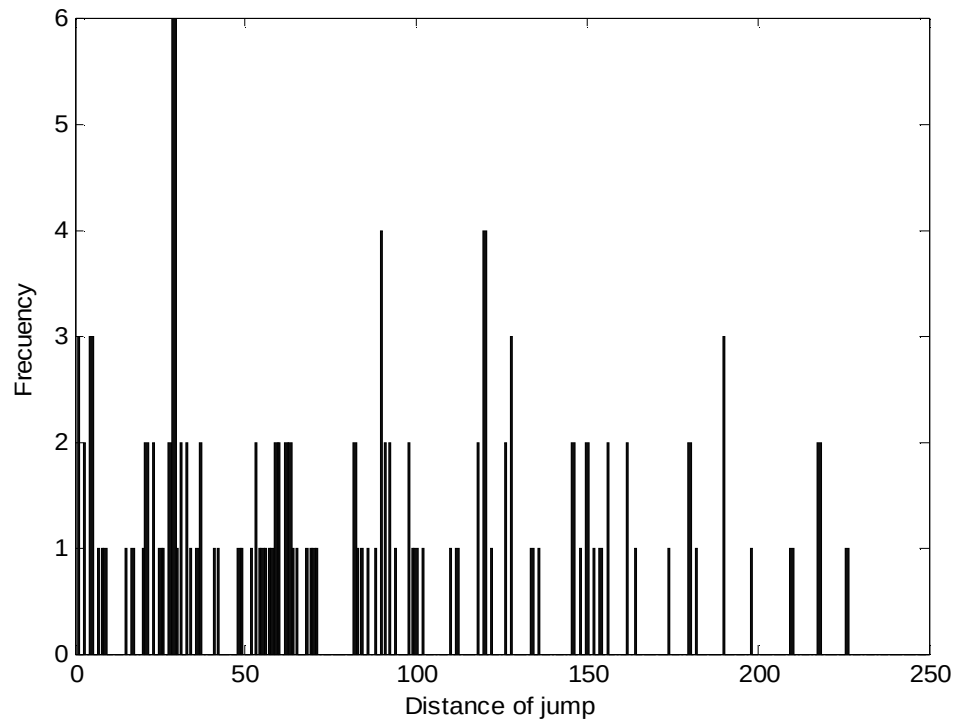

**Fig. 8.** Interleaving distance versus frequency for cdma2000, packet size = 250

## 5. Conclusions

- In essence the principle of interleaving used in cdma2000 is based on manipulating a counter whose value defines the mapping position at the output. Instead, the turbo interleaver used by the W-CDMA standard is based on a permutation method using prime numbers for generating permutations for both rows and columns in a rectangular matrix.

- While output vectors appear to be randomly distributed, the transformation which defines the mapping between input and output positions in both W-CDMA and cdma2000 turbo interleavers is bijective and of deterministic nature, i.e. mapping between input and output is fixed.

- Considering the interleaving average distance and the variance of interleaving distance (Table 18) it can be observed that the W-CDMA and cdma2000 turbo interleavers have a very similar behavior even when their interleaving algorithms are substantially different.

- The distribution of distances, as can be seen at the histograms of Fig. 7 and Fig. 8, corroborate that the dispersion patterns of the W-CDMA and cdma2000 turbo interleavers are statistically quite similar.

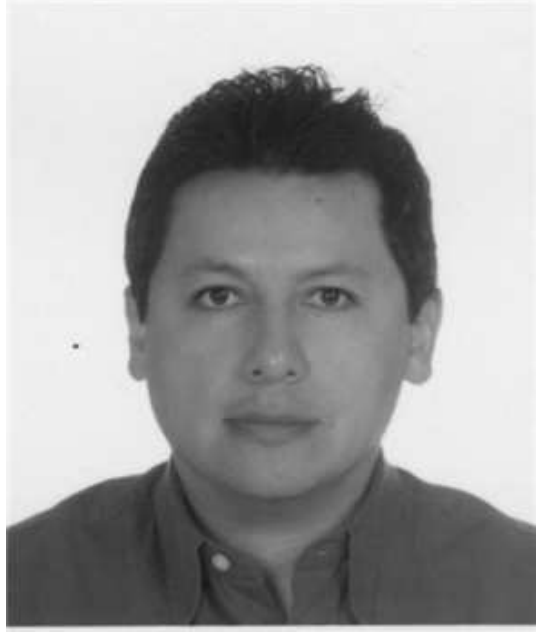

**Fabio G. Guerrero**, received a B.Eng. degree in telecommunications engineering from Universidad del Cauca, Popayan, Colombia (South America), 1992, and a M.Sc. degree in Real-Time Electronic Systems from Bradford University, UK, 1995. Currently, he works as telecommunications assistant lecturer in the Department of Electrical and Electronics Engineering of Universidad del Valle, Cali, Colombia (South America). His research interests include digital communications, telecommunication systems modeling, and next generation networks. He is member of the Communications Society of the IEEE and has served as reviewer for several international journals.

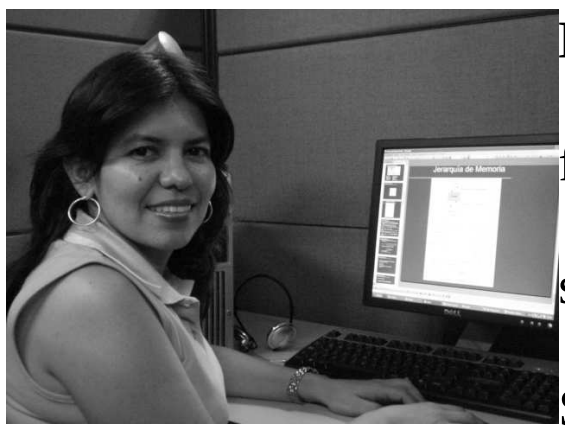

**Maribell Sacanamboy**, She received his 5-year Diploma in Electronics Engineering in 1999 from the Universidad Javeriana, Cali, Colombia (South America). Currently she is a M.Sc. student at the Electrical and Electronics Engineering School of the University of Valle, Cali. She is with the Department of Computer Science Engineering, Universidad Javeriana, Cali. Her research interests include digital design for communication systems, fuzzy logic, and computer architectures.